%% file: main.tex
\documentclass[preprint,12pt]{elsarticle}

\usepackage{subfigure}
\usepackage{amsmath}
\usepackage{amssymb}
\usepackage{tabularx}
\usepackage{adjustbox}
\usepackage{array}
\usepackage{pgfplots}
\usepackage{times}
\usepackage{multirow}
\usepackage{cancel}
\usepackage{tikz}
\usetikzlibrary{positioning,calc,decorations.pathreplacing}
\pgfplotsset{compat=1.5}
\usepackage{epic,eepic}

\usepackage{mathtools}

\DeclarePairedDelimiter\floor{\lfloor}{\rfloor}

\usepackage[boxruled,linesnumbered]{algorithm2e}

\usepackage{lineno}

\journal{arXiv.org}

\begin{document}

\begin{frontmatter}


\title{Max-min Fairness Based Faucet Design for Blockchains}



\author[1]{Serdar Metin\corref{cor1}}
\ead{serdar.metin@boun.edu.tr}

\author[2]{Can \"{O}zturan}
\ead{ozturaca@boun.edu.tr}

\cortext[cor1]{Corresponding author}

\address{Bo\u{g}azi\c{c}i University, Bebek, \.{I}stanbul, Turkey}

\begin{abstract}
In order to have transactions executed and recorded on blockchains such as the Ethereum Mainnet, fees expressed in crypto-currency units of the blockchain must be paid. One can buy crypto-currency called Ether of the Ethereum blockchain from exchanges and pay for the transaction fees. In the case of test networks (such as Rinkeby) or scientific research blockchains (such as Bloxberg), free crypto-currency, Ether, is distributed to users via faucets. Since transaction slots on the blocks, storage and smart contract executions are consuming blockchain resources, Ethers are distributed by fixed small amounts to users. Users may have different amount of Ether requirements; some small amounts and some large amounts during different times. As a result, rather than allowing the user to get a fixed small amount of Ether, a more general distribution mechanism that allows a user to demand and claim arbitrary amounts of Ether, while satisfying fairness among users, is needed. For this end, Max-min Fairness based schemes have been used in centralized settings. Our work contributes a Max-min Fairness based algorithm and its Solidity smart contract implementation that requires low transaction costs independent of the number of users. This is important on the Ethereum blockchain, since a smart contract execution with transaction costs depending on the number of users would mean block gas limit exhaustion problem will eventually be met, making the smart contract ineffective. We report tests which confirm that the low transaction cost aims have been achieved by our algorithm.
\end{abstract}


\end{frontmatter}


\section{Introduction}
\label{intro}

Since its conception in 2008 with Bitcoin \cite{nakamoto2008bitcoin}, blockchain technologies have been the focus of much attention. 
Although successful at achieving its initially proposed purpose of providing a peer-to-peer electronic cash system, Bitcoin was conceived  as an autonomous global currency with guaranteed scarcity, and as such, offered limited programmability and functionality. 
The designers of the following generation of blockchain systems mainly problematised this point and endeavoured on expanding blockchain capabilities. With the introduction of smart contracts by the Ethereum~\cite{wood2014ethereum}, the blockchain technology met Turing-complete programming functionalities.

Our work aims to address a recurrent question in computer science, within the blockchain context: the fair allocation of shared resources. We focus on the fair allocation of intrinsic resources of blockchains. Since a blockchain is a distributed ledger, operated in a distributed manner, the ability to operate on the blockchain (e.g. executing a transaction or a smart contract function, or deploying a smart contract) is a shared, limited resource. We look at the fair allocation of this resource.

In the Ethereum blockchain ecosystem that offers smart contract functionality, the resource
usage mentioned above are quoted
in terms of \textit{gas}, which refers to the cost necessary to perform a transaction on the blockchain.
The gas is priced using the blockchain's intrinsic crypto-currency,  called Ether in the case of Ethereum.  
Hence, just like a number of litres of petrol (priced as USD per litre) is needed in order to have a car 
travel a number of kilometers,  a number of gas units (priced as Ether per gas unit)  is needed to  
execute a number of  instructions in a blockchain transaction.  Thus, the problem collapses down to the 
distribution of the system's intrinsic  crypto-currency. 
In commercial public networks like Ethereum Mainnet, the distribution process relies on the competition to create new Ether units  of the blockchain currency, and the trading of the already generated Ethers. However, 
on  test networks such as Rinkeby or scientific research blockchains such as Bloxberg \cite{bloxberg} alternative 
Ether distribution mechanisms are used.

A \textit{faucet} is one such mechanism, which offers free currency to users according to some predefined policy. In general, faucets offer a fixed amount of currency for a given time period or block span. For example, Bloxberg 
blockchain provides 0.2 Ethers via its web based faucet~\cite{bergfaucet}. 
However, this mechanism can be exploited simply by making recurrent requests and accumulating the obtained currency. For this reason, it cannot be accounted for as a fair distribution scheme. Max-min Fairness 
\cite{bertsekas1992data,keshav1997engineering} is a distribution scheme that is widely employed in different contexts where fairness is a system requirement (e.g. cpu scheduling, bandwidth allocation etc.), and it can also be considered for the fair distribution of currency in faucet systems.

On the Ethereum blockchain, the size of each block is  bound by a maximum amount of gas that can be spent per block. 
This upper bound on the gas amount is known as the {\it block gas limit}. 
A contract function will not be able to execute
if its gas cost exceeds the block gas limit. We refer to this problem as the block gas limit exhaustion problem. 
Hence, smart contract functions should be designed and implemented in such a way that their execution does
not consume too much gas which may lead to block gas limit exhaustion problem.  
If gas limit is reached, it will simply  mean the contract function cannot be executed which in turn may mean
the smart contract can no longer operate properly. 

\begin{figure}[ht]
\centering
    \includegraphics[width=\textwidth]{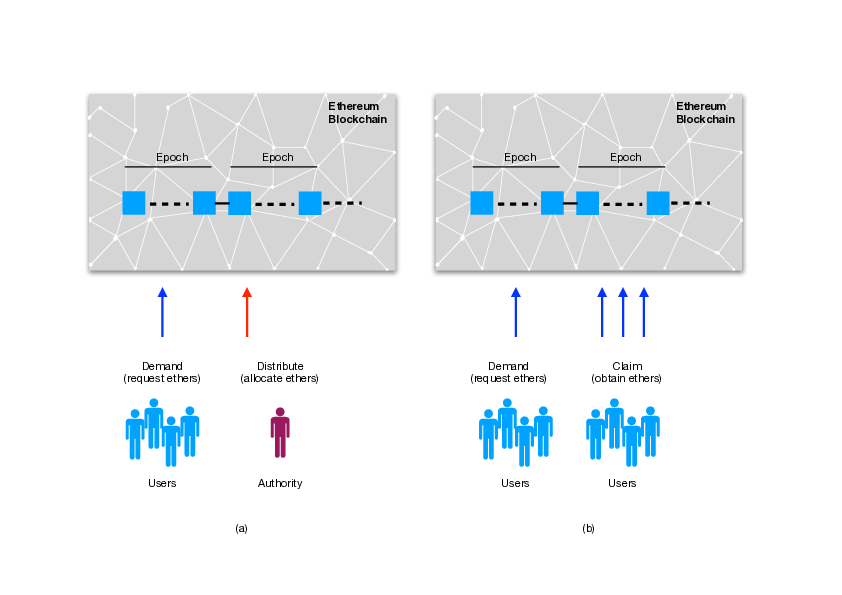}
\caption{The call sequences of the (a) Authority driven faucet, and (b) Autonomous faucet (driven by crowds of users), in a distribution cycle. Blue arrows indicate low-cost transactions (demand and claim), and red arrow indicates a high-cost transaction (distribute).}
\label{lab-fig-overview}
\end{figure}

In this work, we first implement Max-min Fairness algorithm in the blockchain context as a smart contract, as it is originally implemented in centralized systems. After demonstrating the shortcomings of this implementation in the blockchain context (i.e. block gas limit exhaustion problem), we contribute an algorithm that actuates the Max-min Fairness scheme in the blockchain context without running into the original implementation's shortcomings. We name our algorithm Autonomous Max-min Fairness (AMF), since it is operated autonomously by the users in the system, as opposed to the original algorithm, in which the distribution operation is done centrally by an authority. 
Figure~\ref{lab-fig-overview} illustrates the operations of centralised and authority driven faucet smart 
contract implementation in (a)   and  autonomous and decentralised implementation driven  by crowds of users in (b). In the former scenario, the distribution is done with a single call to the \textit{distribute} function by the authority node in the beginning of the epoch, whereas in the latter the users make multiple calls to the \textit{claim} function throughout the epoch in order to obtain their own share.

We further extend the study to a weighted version of Max-min Fairness scheme, in which case the users are assigned weights for their respective shares, according to some prioritisation policy. In the tests we run, the weight of each user's share is defined to be the reciprocal of the total demand volume of the user, up to and including the then present demand. By discouraging unnecessary demands, this policy leads to higher sharing incentives among users. Moreover, it secures fairness of distribution in the long run, since latter allocations are mediated with the former demands of a given user. We name this algorithm Weighted Autonomous Max-min Fairness (WAMF).

The remainder of the article is organised as follows: In the next section, we review the literature on related work. Having laid out the background on blockchain studies, in Section \ref{statement} we state our problem in the light of the observations from Section \ref{related_work}. We continue with a section where we justify our design decisions concerning the experimentation environment. Following that, in Section \ref{max-min} the Max-min Fairness scheme and its adaptation to the present context is explained. In Section~\ref{implementation}, the implementation details are laid out. Once the implementation is explained, the results of the experiments are given in Section~\ref{results}. We discuss the results in 
Section~\ref{discussion}, and conclude the article in Section \ref{conclusion} with the prospects of possible follow-up studies.

\section{Related Work}
\label{related_work}

Blockchain technologies have been proposed for a number of user applications (e.g. \cite{crosby2016blockchain,foroglou2015further,kuo2017blockchain}), and also for background services (e.g. \cite{maesa2017blockchain,samaniego2018virtual}).

By the introduction of tokenised economies, blockchain systems are rendered capable of governing allocation and trade of resources \cite{shirole2020cryptocurrency}. A token is a data structure with certain attributes and operations defined on them, serving for representing items or value. The first two standards that are developed for token economies are ERC$20$ and ERC$721$, which define divisible and non-divisible, or as they are so called, \textit{fungible} and \textit{non-fungible} tokens, respectively. Although compatible with our setting, for reasons of simplicity we did not use token standards in our implementation.

In many areas in computer science where the problem of distributing shared resources is encountered, Max-min Fairness  \cite{bertsekas1992data,keshav1997engineering} has been considered a fair method \cite{hahne1991round,gogulan2012max}. It is also the main method employed in the present study.

The question of fair sharing first arose in the context of operating systems, where scheduling the resources of a single computer (e.g. processor time) among \textit{processes} was the main problem \cite{waldspurger1995lottery}; followed by the problem of distributing the same resources among \textit{users} \cite{kay1988fair}, typically at the computer centres of universities. Similar problems are addressed in the computer networks literature over the allocation of link bandwidth \cite{nace2008max,hahne1991round}. Fair scheduling algorithms have also been the focus of attention in grids~\cite{doulamis2007fair}.

With the advancements in distributed systems, and new paradigms in cluster and high-performance computing, the problem of fairness evolved yet to larger scales, and new questions arose. In this context, typically, service providers charge users for the common resource that is demanded by, and allocated to them. The same question is now expressed in terms of charging fairness: how much should each demand cost, for it to be fair among clients~\cite{marbach2002priority} ? Should each type of resource cost the same, and if not how are they traded~\cite{ghodsi2011dominant} ?   

\section{Problem Statement}
\label{statement}
As indicated in Section \ref{related_work}, the criteria for fair allocation is intimately related with the context the problem is situated in. A number of observations that stand out to be relevant are as follows:

\subsection{Abstraction for Demands: The level at which distribution is done} As observed in Section \ref{related_work}, the allocation may be done at process-level \cite{waldspurger1995lottery}, user-level \cite{kay1988fair}, or group-level \cite{mohanty2020qos}. The faucet system that we contribute is designed so as to provide fair distribution of internal currency among users, with the assumption that the users are identified and registered for making demands. Therefore, the abstraction for the demands are at the \textit{user level}.

\subsection{Abstraction for Resources: The number of resource types and their relationship with each other} If the resources are abstracted to be homogeneous, the main problem is pricing a unit resource \cite{marbach2002priority}. On the other hand, if more than one type of resources abstracted, the problem should be extended as to address how they are traded among each other \cite{ghodsi2011dominant}.

A key factor for determining the price of a cryptocurrency is its by then present and ultimate total supply \cite{hayes2015factors}. In the system we developed, the growth of the total supply is predefined by an immutable policy, securing the ground for a calculable and predictable pricing mechanism. Since the domain we restricted the present study into is non-commercial blockchain systems, pricing here does not refer to monetary pricing, but rather the cost of operations in terms of intrinsic cryptocurrency.

For simplicity, abstraction for the resource is kept at single resource type, which is the intrinsic currency of the blockchain ecosystem, which in our case is the Ether cryptocurrency. With Ether, a user can pay for storage  as well as  execution cost of  smart contract functions on the Ethereum blockchain, providing the basis for a unified resource type.

\subsection{Temporal Granularity: The frequency at which the allocation procedure takes place} For addressing this issue, the blockchain is divided into epochs, which we define as a constant number of successive blocks. According to the original Max-min Fairness algorithm, the distribution should be done at the beginning of each epoch, which we implemented as such. In the algorithms we develop, we also expand the allocation procedure over the whole span of the epoch, to be decentrally carried out by users, in the so called \textit{claim rounds}. The temporal setting will further be described in Sections \ref{implementation} and \ref{results}.

\section{Design Decisions on the Experimentation Environment}
\label{design}

Although there are a number of design decisions for setting up a blockchain system to carry out experiments on, one key factor is the proof scheme employed in the consensus protocol. In the present study, the experiments are carried out on the Parity implementation of a permissioned Ethereum blockchain \cite{parity}. The main concern for this choice is to decouple two independent, yet intertwined questions specific to the blockchain environments. Blockchains are a means both for decentralisation, and for securing digital trust. By decoupling these two questions and allowing to concentrate on decentralisation premise, permissioned blockchains are ideal experimentation environments for blockchain operation analysis. Let us elaborate on that.

Various proof schemes, employed in blockchains' consensus protocols, necessitate different trust assumptions, and equivalently, offer different levels of digital trust to the ecosystems they are embedded in. If one imagines proof schemes on a scale for their provision of trust, Proof-of-Work (PoW) blockchains reside on one extreme, since they assume no preliminary digital trust, and provide all the digital trust needed via their operation. For this reason they are referred to as \textit{trustless computation} environments, in the sense that no prior trust is needed among the users to be involved in the operation of the ecosystem. Initial blockchains such as Bitcoin \cite{nakamoto2008bitcoin} and Ethereum \cite{wood2014ethereum} employ PoW based consensus protocols.

Although they can operate stand-alone trust-wise, PoW blockchains expend enormous physical resources (e.g. electricity, processing power) and their operation is costly. Other proof schemes were proposed to replace PoW in order to eliminate these costs. These schemes provided different levels of trust, compensated by extra-digital measures to different extents, inversely proportional with the former. Among these are: Proof-of-Space \cite{ren2016proof}, Proof-of-Stake \cite{kiayias2017ouroboros,king2012ppcoin}, Proof-of-Prestige \cite{krol2019proof}, Proof-of-Activity \cite{bentov2014proof}, Proof-of-Useful-Work \cite{ball2017proofs}, with different utilities and limitations they bear.

Proof-of-Authority (PoA) blockchains, residing on the other extreme, provide no trust via their operation, and rely solely on extra-digital measures (e.g. reserve the right to operate on the blockchain only to trusted parties) to secure trust.

The trust structure described above for PoA is equivalent to the trust structure of the conventional computation environments, which is referred to as Pretty Good Privacy (PGP) trust chain \cite{abdul1997pgp}. The PGP scheme secures trust with the assumption of the presence of a \textit{trust anchor}, a party that can be unconditionally trusted, and from that point, other parties are trusted either by the direct reference of, or by a chain of references rooted at the trust anchor. In the PoA setting, authority nodes act as trust anchors.

We implemented our algorithms in Solidity programming language and run on an Ethereum Virtual Machine (EVM) environment \cite{dannen2017introducing}, and more specifically, its Parity implementation, as mentioned above. The main reason for selecting this framework is its wide use among blockchain ecosystems. Many blockchain ecosystems and blockchain based systems utilise either EVM or virtual machines similar to EVM, and support Solidity programming language for smart contracts (e.g. \cite{baird2018hedera,tron,cheng2019ekiden,niloy2021blockchain} etc.), and for this reason there are also studies available on the performance \cite{wohrer2018design}, security \cite{wohrer2018smart}, and inspection \cite{bragagnolo2018smartinspect} of the programming language. Not only is it a widespread programming language, Solidity is also Turing complete \cite{wood2014ethereum}, which makes it well suited for general purpose computations. It is a high-level, easy to read, object oriented script language.

A number of smaller design decisions are taken concerning the parameters of Parity Ethereum Virtual Machine and the procedure of the experiments, which is left to be discussed in Section \ref{results}, since their explanation relies occasionally on the implementation of the algorithms we present.

\section{Max-min Fairness Model}
\label{max-min}

The main objective of the Max-min Fairness scheme is to maximise the minimum share given to a user, and its mechanism is based on a trivial fairness scheme, where resources are uniformly distributed among the demanders, each one of the $n$ demanders obtaining $\frac{1}{n}$ of the resource. Max-min Fairness improves the trivial scheme on the premise that not every demander would demand as much as the share that is reserved for them. Accordingly, the Max-min Fairness allocation algorithm takes recursive iterations over the list of demanders, reallocating unused shares of the underdemanders among the overdemanders.

In the first iteration, starting with the smallest demand and proceeding in the ascending order, the algorithm allocates the demanders the minimum of $\frac{1}{n}$ of the capacity ($c$) and their demands (i.e. $\min\{\frac{c}{n}, d_u\}$). At the end of the first iteration, some demanders are fully supplied, and some capacity is left over. The algorithm, in turn, proceeds with updated $n'$ and $c'$, until either all demands are fully supplied, or the capacity is depleted.

\input{max-min_diagram}

The operation of the scheme can be seen in Figure \ref{max-min_diagram}, and its pseudo-code in Algorithm \ref{max-min_pseudo}. In the pseudo-code the demand heaps are denoted by $D_0$ and $D_1$, and individual demands in these heaps are represented by lower case letters, subscripted with $u$, for user id number (i.e. $d_u$).

The balances of users are kept in a vector, and the balance of user $u$ is represented with $b_u$. At each iteration, the maximum available amount to be allocated to each user is recalculated by dividing the remaining capacity by the number of remaining demands, and denoted by $s$, representing the \textit{unit share}.

To illustrate the operation of the algorithm we may consider the following example: Suppose that a resource of $30$ units will be shared among three users, with the demands expressed as $<$$4,11,15$$>$. The algorithm distributes the resource in $3$ iterations. The rounds and the shares assigned in each round can be seen in Table \ref{max-min_example}.

\input{max-min_example}

How the unsatisfied demand, or the leftover capacity will be treated after a distribution period is a decision of policy. In our current work, we implement a policy that discards all the unsatisfied demands, in the case of capacity depletion, and transfers the leftover capacity to the next distribution period, in the case of satisfying all the demands.

The amount that is reserved for each epoch is denoted by $C$. We call this amount the \textit{epoch capacity}, and in the present study, we took it to be constant. The actual amount that is distributed in an epoch is denoted by $c$, and it is at least as much as $C$, since it is added to $c$ at the beginning of each epoch (i.e. Algorithm \ref{max-min_pseudo} line $2$).

In Algorithm \ref{max-min_pseudo}, the lines $4-20$ constitute the main, or outer loop of the algorithm, which is responsible for repeating the inner loop (lines $10-18$) until either the demands or the capacity is depleted. It starts with calculating the share (lines $5-9$), and then starts the inner loop. Once the proceeding of the inner loop is completed, the demand heaps exchange their functions (line $19$) and the outer loop takes another iteration.

The inner loop accounts for iterating on and processing the demands in the active heap. In line $11$ the demand volume and the user id at the root of the heap is read into a variable and deleted from the heap. After that the minimum of user demand and unit share (i.e. $\min\{\frac{c}{n}, d_u\}$) is assigned to the user in lines $12-14$. The control structure in lines $15-17$ checks whether the demand is fully satisfied. If not, the leftover demand is inserted to the heap with the user's id (line $16$) to be processed in further iterations.

Another version of Max-min Fairness is \textit{weighted Max-min Fairness}, in which case the users are weighted over some predefined policy, and the shares are calculated with the weights assigned to each user, individually. In this version, instead of the number of demands, the total capacity is divided by the \textit{total weight} in order to calculate the unit share ($s$). In turn, the \textit{user share} ($s_u$ for user $u$) is calculated for each user by multiplying the unit share with the user's weight. The users are allocated the minimum of their demands, and their individually assigned user shares.

\begin{table}[t]
    \centering
    \input{mf_symbols}
\caption{Symbols used in CMF (Algorithm \ref{max-min_pseudo})  and their meanings}
\label{tab:symbols}
\end{table}

\input{max-min_pseudo}
\\ \\ \\ \\
Accordingly, the formula for calculating the unit share $s$ is:

\[s = \frac{c}{\sum_{j=1}^{n} w_j}\]
\noindent
and the user share $s_u$ is given by:

\[s_u = w_u \cdot s = w_u \cdot \frac{c}{\sum_{j=1}^{n} w_j}\]

We develop autonomous algorithms called  AMF (unweighted version) and  WAMF (weighted version) for actuating  the Max-min Fairness scheme. In WAMF, the weights are defined to be the reciprocals of the total amount of demands users have made up to the distribution time. This aims at incentivizing users to make minimal demands suitable to their needs, in order not to be disadvantageous in the long run. The implementation details of WAMF algorithm, as well as its pseudo-code is presented in Section~\ref{implementation}.

\section{Implementation}
\label{implementation}

The conventional setting to utilise Max-min Fairness typically includes a central unit (either an individual process running on a central processor or a dedicated administrative host in a computer network) calculating the shares and carrying out the iterative assignments. This is applicable to the blockchain context, but not without potential drawbacks. The main bottleneck in such an adaptation is the block gas limit, which imposes an absolute upper bound for the number of operations that may take place within the processing of a single block. For this reason, we implemented two algorithms and compared them. The implementations are available at \cite{metin}

The first algorithm is the \textit{Conventional Max-min Fairness (CMF)}. This algorithm is implemented as if it operates in the conventional computational setting. The demands are collected for a given time period or block span, which is referred to as an \textit{epoch} in this study. At the beginning of the following epoch these demands are supplied resources in the Max-min Fairness order by a single node (typically an authority node) in one step with the \textit{distribute} function.

In the second algorithm, the demands are collected in a given epoch, and the demanders claim their reserved share by calling a \textit{claim} function in the \textit{claim rounds} of the following epoch. We call this approach \textit{Autonomous Max-min Fairness (AMF)}, since there is no need for a central node to carry out the execution, and the system is operated autonomously by its users. The operation of AMF emulates the original algorithm identically, except for the last iteration where the distribution is in the \textit{first come first served} order among overdemanders. Originally, the last iteration is in the \textit{ascending} order of demand volumes, as are all the preceding iterations.

We implemented both unweighted as well as the weighted versions of Max-min Fairness for the AMF. The reason for not implementing a weighted version of CMF is due to its gas cost structure (elaborated on in Section \ref{cmf_results}). In the following subsections we give the implementation details of the algorithms.

\subsection{Conventional Max-min Fairness}
\label{cmf}

As it is in the conventional setting, CMF utilizes two min-heaps, exchanging the demands among each other in each iteration. The operation scheme and the pseudo-code is the same as it is described in Section \ref{max-min} (i.e. Figure \ref{max-min_diagram} and Algorithm \ref{max-min_pseudo}).

Since Solidity does not offer a built-in data structure for min-heaps, we implemented it during  the development of CMF. We kept the implementation of the min-heap minimal in order to keep the gas cost at minimal. Only the amount of demand, and the id (i.e. unique user number given to each user) of the demanding user is stored and operated on. The remainder of the user attributes are fetched from other data structures when needed (e.g. while writing to user balance), by using the user id as the key.

We used an array implementation of heap, a complete binary tree, where the values are kept in a node array and the \textit{insert} and \textit{delete minimum} functions are  implemented so that they index and move the nodes according to the min-heap organisation. This is also immune to degeneration attacks, in which case an attacker feeds the tree with selective input to make one branch grow disproportionately, forcing heap functions run in $\mathcal{O}(n)$ instead of $\mathcal{O}(log(n))$ time.

We present the performance of CMF, as well as the min-heap, in Section \ref{cmf_results}.

\subsection{Autonomous Max-min Fairness}
\label{amf}

In AMF, the epochs are divided into claim rounds. At the end of each round, the remaining number of demands, the remaining capacity, and the resulting share is recalculated. The rounds proceed in this manner until either the capacity is depleted, or all demands are supplied. The rounds are used to emulate the iterations of the outer loop (lines $4-20$ of Algorithm \ref{max-min_pseudo}) of the distribute function.

In order to avoid repetition, we give the pseudo-code only for the weighted version (WAMF), since it is more general as compared to the unweighted version (AMF), the latter being the same algorithm with fewer steps. The pseudo-code of WAMF is presented in Algorithm \ref{wamf_pseudo}. The symbols for the additional variables, and their meanings are given in Table \ref{amf_symbols}. The calculation of weights is obscured from the pseudo-code for the ease of review, and the weights are simply shown as constant variables. The calculation of weights is described in detail in the next subsection.

In AMF, instead of a single-handedly operating \textit{distribute} function, there is a \textit{claim} function, which after necessary checks, allows the user assign her allocated share to herself. Each user is expected to execute the function individually, to have carried out the iterations of the inner loop of the \textit{distribute} function (lines $10-18$ of Algorithm \ref{max-min_pseudo}), in a decentralized manner.

Any share unclaimed in its due round/epoch is lost. It is included in the following round/epoch as part of the leftover capacity. In a given epoch, users may make new demands for the next epoch, while claiming their share for the previous. The time frame can be traced in Table \ref{amf_example} over the demands and corresponding claims, and can be seen more explicitly in Figure \ref{cascade}.

\input{amf_example}

\input{epoch}

\begin{table}[ht]
    \centering
    \scalebox{0.85}{
        \input{amf_symbols}
    }
\caption{Symbols used in Algorithm \ref{wamf_pseudo} and their meanings}
\label{amf_symbols}
\end{table}

In AMF the demands are kept in a map, rather than a min-heap, since it is necessary for each user to be able to access their own demand entry, while claiming it. In the present implementation, the demands are kept for one epoch, and claimed in the following. For this reason, a circular buffer of size two is kept for each user, in order to prevent an incoming demand in a given epoch to overwrite the previous epoch's demand, before it is claimed. This leads to a two dimensional ($2\text{ x }n$) demand vector, where the demands for even and odd epochs are kept separately. Additionally, the variable for keeping the epoch in which the demand was made (for preventing an obsolete demand to interfere with later demands) is implemented; likewise as a circular buffer of size two, in order to separate between the even and the odd epochs.

In addition to the restructured \textit{demand}, and the newly introduced \textit{claim} functions, AMF includes a state update function, which is called at the beginning of both. The state update function checks the block number, and calculates the epoch and the round in which the called function will be executed (lines $3$ and $10$, respectively). The number of blocks for the duration of an epoch and a round, is also a parameter of the system, which we experimented on in the present study, and commented on in the results section.

The pseudo-code in Algorithm \ref{wamf_pseudo} is organised in three functions, namely, \textit{update state} (lines $1-15$), \textit{demand} (lines $16-24$), and \textit{claim} (lines $25-45$). At the beginning of each function (in lines $2$, $18$, and $27$) a local selector variable ($i$) for the circular buffers is declared and defined. When called in a given epoch, the state update and the claim functions assume the same selector values, and demand function assumes its binary complement. That is to say $i$ values proceed as $<0,1,0,1,...>$ for the \textit{state update} and \textit{claim} functions, and as $<1,0,1,0,...>$ for the \textit{demand} function.

In line $3$, the epoch number ($E$) is checked for. If the value of $E$ is found to be obsolete, it is updated. Once the epoch number is updated, the round number, the capacity, and the unit share are also updated (lines $5-7$), and the function returns. If epoch number is found to be up-to-date, a similar check is done for the round number in line $10$. This check, when it returns positive, leads to the update of the round number and the unit share (lines $11-12$), and the function returns. If no update is required, the function returns without making any changes in the state.

After updating the state and setting the selector variable, in line $19$ the demand function checks whether the user has made a demand in the then present epoch. If the user has made a demand, the function returns without registering the newly arrived demand. If not, the demand amount ($a$) is written to the corresponding slot in the circular demand buffer of the user, and the demand epoch of the user is updated to be the then current epoch (lines $20-21$). In the following line the function checks whether any demands have been made by other users in the then current epoch. If not, the total weight is set to the user's weight (line $23$), which resets the total weight variable for the next epoch. The variable for keeping the last epoch in which the total weight is reset ($RE$) is updated in line $24$. If demands have been made by other users prior to the then current call (i.e. $RE = E$) the weight of the user is added to the total weight, to be accounted for in the next epoch (line $26$).

The claim function, similar to the demand function, starts with updating the state and initiating the selector variable. It continues with a number of checks (line $33$). Unless the demand has been done in the previous epoch and is greater than $0$, or if the capacity is depleted, the function returns without taking any further action. Following that in line $36$ the function checks whether the user has made any claims in the then current epoch. If so, the last round the user made a claim is checked (line $37$). If that also turns positive, which means the user has claimed her fair share for the round, the function returns without making any assignments.

If the check in line $36$ turns out negative, meaning this is the user's first claim in the then present epoch, the variable for the last epoch the user made a claim ($ce_u$) is updated (line $41$). After that, a similar variable for the round ($cr_u$) is updated in line $41$. Next, the assignment operations similar to the ones in Algorithm \ref{max-min_pseudo} is done in lines $44-46$.

Note that this algorithm differs from the CMF algorithm in that the leftover demands are not inserted into another heap; they remain in the map. Instead, the fully satisfied demands are removed from the cumulative weight variable in lines $42-44$, having the same effect as deleting the minimum in CMF algorithm. This way, as long as there is an unsatisfied demand, the user's weight is included in the total weight, and the unit share is calculated accordingly. At the end of the epoch, all demands are obsoleted.

\input{wamf_pseudo}

\subsection{Weighted Autonomous Max-min Fairness}
\label{wamf}

As the operation of the algorithm is described in Section~\ref{amf}, the only part that is left to be explained in this subsection is the calculation of weights.

We defined weights to be the multiplicative inverses of the total demand volume, up to and including the then present demand. This poses a problem in the smart contract context, since Solidity does not offer floating point data types. In other words, since the demand volumes are defined to be positive integers, it is not possible to keep weights as they are, since the value needs floating point data type to be stored. Instead, we keep the total demand volume for each user ($dt_u$ for user $u$), introduce an intermediary variable $p$ (standing for \textit{precision}) and take the weight equal to:

\[w_u = \floor*{\frac{p}{dt_u}}\]

We get rid of this intermediary variable while calculating the unit share. Therefore, instead of

\[s = \floor*{\frac{c}{\sum_{u=1}^{n} w_u}}\]
\noindent 
we use:

\[s = \floor*{\frac{c \cdot p}{\sum_{u=1}^{n} w_u}}\]
\noindent 
since

\[s = \floor*{\frac{c \cdot p}{\sum_{u=1}^{n} \frac{p}{dt_u}}} = \floor*{\frac{c}{\sum_{u=1}^{n} \frac{1}{dt_u}}} \]
\noindent 
Similarly, while calculating the user share we use the intermediary variable $p$:

\[s_u = \floor*{\frac{s \cdot \floor*{\frac{p}{dt_u}}}{p}}\]
\noindent 
As long as the value of $p$ is larger than the total demand volume of the user, we obtain non-zero weights from $\floor*{\frac{p}{dt_u}}$. For $p = 10^k, k \in \mathcal{Z^+}$ is the number of decimal places stored for weights.

\section{Results}
\label{results}

In this section, we present the results over the gas costs used as the main performance metric. The tests are run on Parity Ethereum $2.7.2$, and the contracts are implemented using Solidity $0.5.13$, thus the gas costs are according to the definitions given thereby.

In our tests, we run Parity in development mode and used its \textit{instant seal} consensus algorithm, in which each transaction is placed in an individual block and inserted instantly to the blockchain. A convenient metric for measuring time is the block number. In the deployment of the system, this metric can be used with the block latency to come up with rough temporal estimations.

Since block latency is a policy parameter for each blockchain ecosystem, taking block number as the main temporal performance metric is convenient also in terms of generalizability of the results. As it is presented here, our results are independent of consensus algorithm, and block latency parameters.

The results for each algorithm are presented in the subsections below. The data are available at \cite{metin}

\subsection{CMF Results}
\label{cmf_results}

As indicated in Section \ref{cmf}, in the CMF, the demand vector is implemented as an array of two min-heaps, exchanging the demands among each other at each iteration. The demands arriving from the users are collected in $D_0$ for the span of an epoch. At the end of the epoch, the distribute function is called by the authority node, and the distribution is done. The first iteration is done over $D_0$, taking all demands from the smallest to the largest, granting the available share to the user, and finally either deleting the minimum demand, if it is completely supplied, or deleting it from $D_0$ and inserting it to $D_1$, otherwise, to be supplied in the next iterations if possible. The heaps exchange functions, and the process is repeated until either all the demands are supplied, or the capacity for the epoch is exhausted (see Algorithm \ref{max-min_pseudo})

Gas usage averages for $n=100$ entry sets are shown in Table \ref{heap_results}. For comparison, the gas performance of a general case heap implementation~\cite{zmitton}, called Eth-heap,  is provided next to our results:

\input{heap_results}

Considering the $8.000.000$ block gas limit, the heap operations impose an upper bound of $60$ entries to be processed per block, on average, as seen with the cost of operations in Table \ref{heap_results}. This number is to be further lowered  with the additional cost of assignment operations, needed to record the fair share of each user to her balance.

The finding immediately implies that an algorithm implemented as a smart contract and relying on a central node to carry out the distribute function, cannot support more than $\sim10$ users, assuming that $3$ iterations are necessary on average for a distribution process to complete. The exact number is a function of how disperse the demands are, since the number of delete/insert operations is dependent on the number of iterations necessary to answer all the demands, which in turn is dependent on how disperse the demands are.

This is also the reason why a weighted version of CMF has not been implemented in the present study. The extra cost of calculating and storing weights will make the weighted version perform even worse than the unweighted version.

\subsection{AMF and WAMF Results}
\label{amf_results}

The first advantage to be pointed out for AMF is that it virtually has no limit for the number of users that the system can support. The average gas costs of \textit{demand} and \textit{claim} functions for a system with $10, 50, 100$ and $500$ users can be seen in Table \ref{amf_performance}. The tests have been carried over in a setting where users have made demands, and claimed their demands in the succeeding epoch. The results indicate that several \textit{demand} and \textit{claim} function calls can be included within a block, without running into the block gas limit exhaustion problem.

\input{amf_performance}

The results also indicate that the cost of \textit{demand} and \textit{claim} functions do not grow with the growing number of users. On the contrary, there is a slight decrease in the average costs, with the growing number of users. The reason for this is the fact that in each epoch the first call to both functions are costlier, since state variables are updated in these calls. With large sample sizes, this difference tends to even out better as compared to the relatively smaller sample sizes.

It should also be noted that the epochs and rounds should last enough for each user to be able to make claims and demands. Since the instant seal engine deployed in the tests place each transaction in an individual block, the epoch and round spans are so chosen as to allow each user be able to make claims and demands within an epoch. The parameters of the system that the tests have been carried on have been shown in Table \ref{amf_parameters}.

According to this, in a setting with $n$ users, in the first epoch, $n$ blocks are used for user registration function calls and $2n$ blocks are filled with empty transactions in order to synchronise the process. The following demand function calls occupied $n$ more blocks, concluding the first epoch. From the second epoch on, the sequence is $3$ rounds of claim in $3n$ blocks, followed by $n$ blocks of demand for the next epoch. Three sets are run (adding up to $4$ epochs), and the averages are collected.

\input{amf_parameters}

One thing that should be accounted for is that the average cost of demand function declines throughout the rounds. The reason for this is, some demands have been fully supplied in the previous epoch, thus, fewer calls to claim function lead to the full execution of the function (i.e. calls from users whose demands have already been satisfied return without making any assignments). The average claim costs of rounds for Max-min and Weighted Max-min Fairness schemes can be seen in Table \ref{claim}.

\input{claim}

The number of rounds, as indicated in Section \ref{cmf_results} is a function of the initial distribution of the demands. In our tests, we drew random demands from an approximately uniform distribution offered by Javascript Math.random() function, in the range $[10,30)$, and the epoch capacity is set to $20n$, so that on average the overdemands and underdemands could balance each other out.

In all the simulations with a Python script, the distribution is completed in $3$ iterations. Therefore, in the tests presented here, we run the system for $3$ rounds of claims. The results are cross-checked with the Python simulations and proved identical. We suspect that with the parameters used in this study, $3$ iterations might be an upper bound, but we do not have a proof. Further investigation needs to be carried out to in order to come up with a theoretical
bound.

Another variable that can be parameterized according to the policy and that would effect gas costs is the size of the variables used to represent amounts. The size of the variables can be chosen smaller to save from the extra cost of unused space. The necessary sizes for the variables is dependent on the total amount that is planned to be distributed in the long run, maximum available allocation in an epoch and  the maximum number of epochs to distribute all the resource etc. In the present study, all the variables are implemented as their $256$ bit defaults, in order not to lose generality.

\section{Discussion}
\label{discussion}

The main bottleneck, and the main performance metric of the present study is the gas consumption, and this is arguably a natural approach for studies on blockchain systems. However, the results presented in this study are not to be taken for their absolute values. Low level improvements may be introduced in coding or compilation, leading to lower transaction costs. The aim of this approach is to demonstrate the availability, and the cost \textit{structure} of the Max-min Fairness algorithm, and its different implementations.

Accordingly, the present study demonstrates, over the failure of CMF to support more than 10 users, that it is not feasible for Max-min Fairness scheme to be implemented in the blockchain context as it is implemented in the conventional computational settings. In principle, because of the block gas limit, blockchain systems are not well suited for algorithms, which cannot be efficiently distributed to be processed by multiple computing parties, with partial data, and asynchronously. A single transaction to carry out a function with heavy computational burden is not a working strategy while developing software for blockchain systems.

This is in accordance with the distributed nature and the philosophy of the blockchain systems. In contrast with the centralized systems, blockchains aim to distribute both the work and the control among its users. For this reason, they are \textit{incentive driven}, as opposed to centralized systems, which are \textit{authority driven}. That is to say, centralized systems rely on an authorized component (e.g. operating system kernels, load balancers, web servers etc.) to carry out the computation; whereas blockchain systems rely on incentivising its users to operate the system in a way that the outcome will turn out to be the desired computation. Both AMF and WAMF are designed taking those points into consideration. Consequently, they offer scalable solutions for blockchain systems.

Another possibility to consider is changing the capacity replenishment policy. In the present study, the capacity is replenished by a constant quantity $C$ at the beginning of each epoch. Instead, the tests can be run with varying quantities of replenishment over time, possibly according to some function of epoch number (i.e. $C = f(E)$). This may serve as a distribution mechanism for systems that run on donations, like election rallies or other types of fund raising projects, where public transparency, responsibility, incentivisation, and participation are matters of consideration. This kind of a distribution mechanism lends these projects the opportunity to be publicly transparent, and make commitments (e.g. declaring the weights for the expenditure items) prior to raising funds, since the system assures the enforcement of declared commitments, by the virtue of its immutability.

\section{Conclusion}
\label{conclusion}

In the present study we addressed the problem of fair distribution of shared resources within the blockchain systems context. We worked on the intrinsic resources of blockchains, and developed faucets as smart contracts, running different implementations of Max-min Fairness Algorithm, which is traditionally accepted realizing fairness in the literature.

It has been demonstrated that the Max-min Fairness algorithm, as it is implemented in the conventional programming contexts, cannot support a public system because of the scaling of its gas cost structure. Two autonomous implementations of the algorithm are offered as a solution, and the tests have shown that these implementations can support wide public use of the system without running into block gas limit exhaustion problem.

Although, in the present, the faucets are mainly utilised as tools for distributing the native currency of the test networks, the operation of faucet systems need not be limited to this use case. These systems have the ability to represent any resource type, and accordingly, to fairly allocate them, as briefly discussed in the preceding section.

The faucet algorithms presented in this study are designed for single resource planning. For the prospective studies we might propose focusing on multi-resource planning problems. One way would be keeping each resource type separate and distribute them in parallel, with the algorithms developed in the present study. This will assume an independence of the supply and the value of the resources. If the resources are dependent on each other, alternative solutions must be proposed and evaluated.

\bibliographystyle{elsarticle-num-names}
\bibliography{bibl}
\end{document}

%% file: max-min_diagram.tex
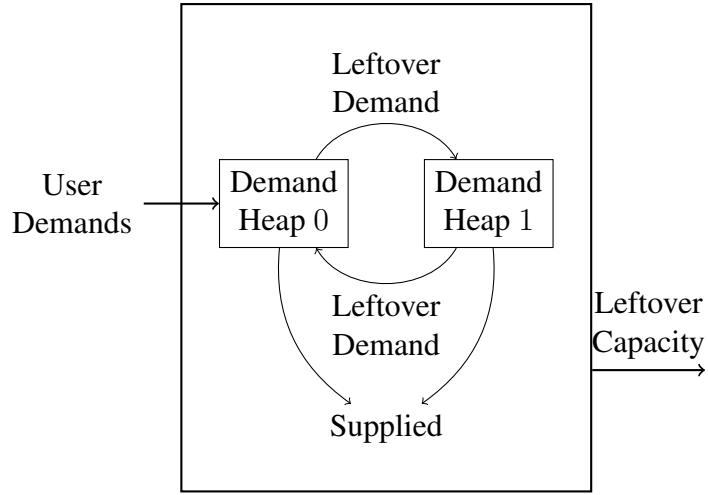
\begin{figure}
    \centering
    \begin{tikzpicture}
        \node [draw,rectangle,align=center] (dho) {Demand\\Heap $0$};
        \node [draw,rectangle,right = of dho,align=center] (dhi) {Demand\\Heap $1$};
        \node (sup) at ($(dho)!.5!(dhi)+(0,-3)$) {Supplied};
        
        \draw [->] ($(dho.north)!.5!(dho.north east)$) to [bend left, above, out=60, in=120] node [align=center] {Leftover\\Demand} ($(dhi.north west)!.5!(dhi.north)$);
        \draw [<-] ($(dho.south)!.5!(dho.south east)$) to [bend left, below, out=-60, in=-120] node [align=center] {Leftover\\Demand} ($(dhi.south west)!.5!(dhi.south)$);
        
        \draw [->] (dho) to [bend right] (sup);
        \draw [->] (dhi) to [bend left] (sup);
        
        \draw[thick] ($(current bounding box.south west) - (.5,.5)$) rectangle ($(current bounding box.north east) + (.5,.5)$);
        
        \draw [<-,thick] (dho.west) --++ (-1,0) node [left,align=center] {User\\Demands};
        \draw [->,thick] ($(current bounding box.east)!.5!(current bounding box.south east)$) --++ (1.5,0) node [midway,above,align=center] {Leftover\\Capacity};
    \end{tikzpicture}
    \caption{The operation of Max-min Fairness Algorithm}
    \label{max-min_diagram}
\end{figure}

%% file: max-min_example.tex
\begin{table}
    \centering
    \begin{tabular}{r|rrr|rr}
         \multicolumn{1}{r}{}  & User $1$   & User $2$   & User $3$  & Share    & Capacity  \\
         \hline\hline
         Demands        & $4$   & $11$  & $15$  &           & $30$      \\
         \hline
         Iteration $1$  & $4$   & $10$  & $10$  & $10$      & $6$       \\
         \hline
         Iteration $2$  & $0$   & $1$   & $3$   & $3$       & $2$       \\
         \hline
         Iteration $3$  & $0$   & $0$   & $2$   & $2$       & $0$       \\
         \hline
         Total          & $4$   & $11$  & $15$  &           &           \\
         \hline
    \end{tabular}
    \caption{An exemplary distribution according to Max-min Fairness scheme}
    \label{max-min_example}
\end{table}

%% file: mf_symbols.tex
\begin{tabular}{|c|p{2.2in}|} \hline
 Symbol     &   Meaning  \\ \hline \hline
 $C$        &   Amount of resource that is added to the existing capacity at every epoch, $C \in \mathbb{Z}^+$  \\ \hline
 $D_i$        &   Set of demand heaps, $i \in \{0,1\}$  \\ \hline
 $U$        &   Set of users $u \in \{ u_1,\ldots,u_n \}$  \\ \hline
 $c$        &   The existing capacity, initialised at $0$, incremented by $C$ at every epoch \\ \hline  
 $s$        &   Unit share  \\ \hline
 $u$        &   User  $u$, $u \in U $  \\ \hline
 $d_{ui}$   &   Demand of user $u$ stored on heap $D_i$  \\ \hline
 $b_{u}$    &   Resource balance of user $u$, $b_u \in \mathbb{Z}^{\ge 0}, u \in U $  \\ \hline
\end{tabular}

%% file: max-min_pseudo.tex
    \scalebox{0.7}{
        \begin{algorithm}[H]
        \SetAlgoLined
        \caption{Max-min Fairness  (CMF) }
        \label{max-min_pseudo}
        \SetKwFunction{function}{\textsc{Function:}}
        \SetKwFunction{distribute}{\textsc{Distribute}}
        \function \distribute{$D,U,c$}\\
        $c \gets c + C$\;
        $i \gets 0$\;
        
        \While{$D_i.size() > 0$ \textbf{and} $c > 0$}{
           \eIf{$c < D_i.size$}{$s \gets 1$ \;}{$s \gets \floor*{ \frac{c}{D_i.size} } $ \;}
           
           \While{$D_i.size > 0$ \textbf{and} $c > 0$}{
                $(d_{ui},u) \gets  D_i.delMin()$\;
                $a \gets \min{(s, d_{ui})}$\;
                $b_u \gets b_u + a$\;
                $c \gets c - a$\;
                \If{$d_{ui} > s$}{$D_{1-i}.insert(d_{ui} - s, u)$\;}
            }
            $i \gets 1-i$\;
        }
        $\Return$\;
        \end{algorithm}
        }

%% file: amf_example.tex
\begin{table}[ht]
    \centering
    \scalebox{0.75}{
        \begin{tabular}{c|c|c|ccc|cc}
            
            \multicolumn{3}{c}{} & User 1    & User 2    & User 3    & Share & Capacity \\
            \hline \hline
            \multirow{4}{*}{} & \multicolumn{2}{c|}{Demand 1} & 4 & 11 & 15 &  & \\
            \cline{2-8}
               & \multirow{3}{*}{}  & Round 1 &  &   &   &   &   \\
            \cline{3-8}
             Epoch 1 & Claim 0 & Round 2 &  &   &   &   &  \\
            \cline{3-8}
                &                           & Round 3 &  &   &   &   &   \\
            \hline
            \hline
            \multirow{4}{*}{} & \multicolumn{2}{c|}{Demand 2} & 11 & 3 & 8 &  & 30\\
            \cline{2-8}
                & \multirow{3}{*}{}  & Round 1 & 4 & 10 & 10  & 10 & 6 \\
            \cline{3-8}
             Epoch 2 & Claim 1 & Round 2 &   &  1 & 3  & 3  & 2 \\
            \cline{3-8}
                &                           & Round 3 &   &    & 2  & 2  & 0 \\
            \hline
            \hline
            \multirow{4}{*}{} & \multicolumn{2}{c|}{Demand 3} & 7  & 8  & 12  & 10 & 30 \\
            \cline{2-8}
                & \multirow{3}{*}{}  & Round 1 & 10 & 3  & 8  & 10 & 9  \\
            \cline{3-8}
             Epoch 3 & Claim 2 & Round 2 & 1  &    &    & 9  & 8  \\
            \cline{3-8}
                &                           & Round 3 &    &    &    &    &    \\
            \hline
            \hline
            \multirow{4}{*}{} & \multicolumn{2}{c|}{Demand 4} & 17 & 13 & 5  &    & 38 \\
            \cline{2-8}
                & \multirow{3}{*}{}  & Round 1 & 7  & 8  & 12 & 12 & 11  \\
            \cline{3-8}
             Epoch 4 & Claim 3 & Round 2 &    &    &    &    &     \\
            \cline{3-8}
                &                           & Round 3 &    &    &    &    &    \\
            \hline
            \hline
           \multirow{4}{*}{} & \multicolumn{2}{c|}{Demand 5}  & .. & .. & .. & .. & 41 \\
            \cline{2-8}
                & \multirow{3}{*}{}  & Round 1 & 13 & 13 & 5 & 13 & 10  \\
            \cline{3-8}
             Epoch 5 & Claim 4 & Round 2 & 4  &    &     & 10  & 6  \\
            \cline{3-8}
                &                           & Round 3 &    &    &    &    &    \\
            \hline
        \end{tabular}
    }
    \caption{An exemplary distribution carried out with AMF}
    \label{amf_example}
\end{table}

%% file: epoch.tex
\begin{figure}
\begin{adjustbox}{width=0.97\columnwidth,center}
\setlength{\unitlength}{3947sp}%
\begingroup\makeatletter\ifx\SetFigFont\undefined%
\gdef\SetFigFont#1#2#3#4#5{%
  \reset@font\fontsize{#1}{#2pt}%
  \fontfamily{#3}\fontseries{#4}\fontshape{#5}%
  \selectfont}%
\fi\endgroup%
\begin{picture}(6024,2430)(1789,-2101)
\put(4441,-436){\makebox(0,0)[lb]{\smash{{\SetFigFont{12}{14.4}{\familydefault}{\mddefault}{\updefault}{\color[rgb]{0,0,0}Epoch-3}%
}}}}
\thinlines
{\color[rgb]{0,0,0}\put(1801,-661){\line( 1, 0){2400}}
\put(4201,-661){\line( 0,-1){300}}
\put(4201,-961){\line( 1, 0){1200}}
\put(5401,-961){\line( 0,-1){300}}
\put(5401,-1261){\line( 1, 0){1200}}
\put(6601,-1261){\line( 0,-1){300}}
\put(6601,-1561){\line( 1, 0){1200}}
\put(7801,-1561){\line( 0,-1){300}}
\put(7801,-1861){\line(-1, 0){2400}}
\put(5401,-1861){\line( 0, 1){300}}
\put(5401,-1561){\line(-1, 0){1200}}
\put(4201,-1561){\line( 0, 1){300}}
\put(4201,-1261){\line(-1, 0){1200}}
\put(3001,-1261){\line( 0, 1){300}}
\put(3001,-961){\line(-1, 0){1200}}
\put(1801,-961){\line( 0, 1){300}}
}%
{\color[rgb]{0,0,0}\put(3001,-661){\line( 0,-1){300}}
\put(3001,-961){\line( 1, 0){1200}}
\put(4201,-961){\line( 0,-1){300}}
\put(4201,-1261){\line( 1, 0){1200}}
\put(5401,-1261){\line( 0,-1){300}}
\put(5401,-1561){\line( 1, 0){1200}}
}%
{\color[rgb]{0,0,0}\put(4426,239){\line(-1, 0){900}}
}%
{\color[rgb]{0,0,0}\put(5251,239){\vector( 1, 0){1140}}
}%
\put(1951,-886){\makebox(0,0)[lb]{\smash{{\SetFigFont{12}{14.4}{\rmdefault}{\mddefault}{\updefault}{\color[rgb]{0,0,0}Demand-1}%
}}}}
\put(3151,-1186){\makebox(0,0)[lb]{\smash{{\SetFigFont{12}{14.4}{\familydefault}{\mddefault}{\updefault}{\color[rgb]{0,0,0}Demand-2}%
}}}}
\put(3263,-886){\makebox(0,0)[lb]{\smash{{\SetFigFont{12}{14.4}{\rmdefault}{\mddefault}{\updefault}{\color[rgb]{0,0,0}Claim-1}%
}}}}
\put(2041,-436){\makebox(0,0)[lb]{\smash{{\SetFigFont{12}{14.4}{\familydefault}{\mddefault}{\updefault}{\color[rgb]{0,0,0}Epoch-1}%
}}}}
\put(3241,-436){\makebox(0,0)[lb]{\smash{{\SetFigFont{12}{14.4}{\familydefault}{\mddefault}{\updefault}{\color[rgb]{0,0,0}Epoch-2}%
}}}}
\put(5881,-2086){\makebox(0,0)[lb]{\smash{{\SetFigFont{12}{14.4}{\rmdefault}{\mddefault}{\updefault}{\color[rgb]{0,0,0}...}%
}}}}
\put(5663,-1486){\makebox(0,0)[lb]{\smash{{\SetFigFont{12}{14.4}{\familydefault}{\mddefault}{\updefault}{\color[rgb]{0,0,0}Claim-3}%
}}}}
\put(5551,-1786){\makebox(0,0)[lb]{\smash{{\SetFigFont{12}{14.4}{\familydefault}{\mddefault}{\updefault}{\color[rgb]{0,0,0}Demand-4}%
}}}}
\put(5641,-436){\makebox(0,0)[lb]{\smash{{\SetFigFont{12}{14.4}{\familydefault}{\mddefault}{\updefault}{\color[rgb]{0,0,0}Epoch-4}%
}}}}
\put(7073,-2086){\makebox(0,0)[lb]{\smash{{\SetFigFont{12}{14.4}{\rmdefault}{\mddefault}{\updefault}{\color[rgb]{0,0,0}...}%
}}}}
\put(6885,-1786){\makebox(0,0)[lb]{\smash{{\SetFigFont{12}{14.4}{\familydefault}{\mddefault}{\updefault}{\color[rgb]{0,0,0}Claim-4}%
}}}}
\put(6863,-436){\makebox(0,0)[lb]{\smash{{\SetFigFont{12}{14.4}{\familydefault}{\mddefault}{\updefault}{\color[rgb]{0,0,0}Epoch-5}%
}}}}
\put(4463,-1186){\makebox(0,0)[lb]{\smash{{\SetFigFont{12}{14.4}{\familydefault}{\mddefault}{\updefault}{\color[rgb]{0,0,0}Claim-2}%
}}}}
\put(4351,-1486){\makebox(0,0)[lb]{\smash{{\SetFigFont{12}{14.4}{\familydefault}{\mddefault}{\updefault}{\color[rgb]{0,0,0}Demand-3}%
}}}}
\put(4576,164){\makebox(0,0)[lb]{\smash{{\SetFigFont{12}{14.4}{\rmdefault}{\mddefault}{\updefault}{\color[rgb]{0,0,0}Time}%
}}}}
{\color[rgb]{0,0,0}\put(6601,-1561){\line( 0,-1){300}}
}%
\end{picture}%
\end{adjustbox}
\caption{Time frame for the matching \textit{demand} and \textit{claim} function calls}
\label{cascade}
\end{figure}
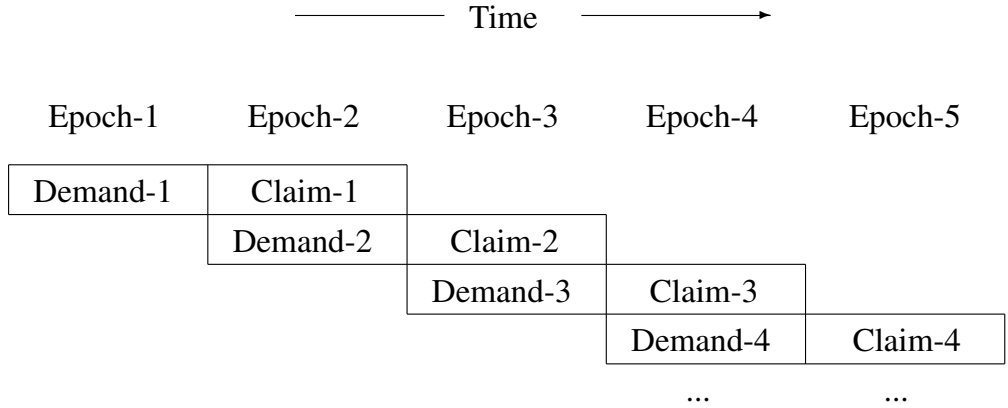

%% file: amf_symbols.tex
        \begin{tabular}{|c|l|} \hline
         Symbol     &   Meaning  \\ \hline \hline
         $C$        &   Amount of resource that is added to the existing capacity at every epoch, $C \in \mathbb{Z}^+ $  \\ \hline
         $B$        &   Current block number  \\ \hline
         $O$        &   The block number at which the contract was deployed, offset \\ \hline
         $E$        &   Epoch number  \\ \hline
         $R$        &   Round number  \\ \hline
         $RE$       &   Reset epoch, the epoch at which the total weight was last reset  \\ \hline
         $ES$       &   Number of blocks in an epoch, epoch span  \\ \hline
         $RS$       &   Number of blocks in a round, round span  \\ \hline
         $U$        &   Set of users $u \in \{ u_1,\ldots,u_n \}$  \\ \hline
         $W_i$      &   Total weight for even and odd epochs, $i \in \{0,1\} $ \\ \hline
         $a$        &   Demand volume, amount   \\ \hline
         $u$        &   User  $u$, $u \in U $ \\ \hline
         $d_{ui}$   &   Demand of user $u$ in list $i$, $i \in \{0,1\}$  \\ \hline
         $de_{ui}$     &   The last epoch user $u$ made a demand, $i \in \{0,1\}$ \\ \hline
         $ce_u$     &   The last epoch user $u$ made a claim \\ \hline
         $cr_u$     &   The last round user $u$ made a claim \\ \hline
         $b_u$      &   Resource balance of user $u$, $b_u \in \mathbb{Z^+} $  \\ \hline
         $w_u$      &   Weight of user $u$  \\ \hline
         $c$        &   The existing capacity, initialised at $0$, incremented by $C$ at every epoch \\   \hline
        \end{tabular}

%% file: wamf_pseudo.tex
\scalebox{0.7}{        
        \begin{algorithm}[H]
        \SetAlgoLined
        \caption{Weighted Autonomous Max-min Fairness (WAMF) }
        \label{wamf_pseudo}
        \SetKwFunction{function}{\textsc{Function:}}
        \SetKwFunction{updateState}{\textsc{Update State}}
        \SetKwFunction{demand}{\textsc{Demand}}
        \SetKwFunction{claim}{\textsc{Claim}}
        \function \updateState{$O, B, E, ES, RS$}\\
        $i \gets E~~mod~~2$\;
        \If{$E < \floor*{ \frac{B - O}{ES} }$}
            {$E \gets \floor*{ \frac{B - O}{ES} }$\;
            $R \gets \floor*{ \frac{(B - O) \% ES}{RS} }$\;
            $c \gets c + C$\;
            $s \gets \floor*{c / W_i}$\;
            \Return
            }
        \If{$R < \floor*{ \frac{(B - O) \% ES}{RS} }$}
            {$R \gets \floor*{ \frac{(B - O) \% ES}{RS} }$\;
            $s \gets c / W_i$\;
            \Return\;}
        \Return\;
        \function \demand{$u, \text{a}$}\\
        \updateState{$O, B, E, ES, RS$}\;
        $i \gets (E + 1)~~mod~~2$\;
        \If{$de_{ui} \neq E$}
            {$d_{ui} \gets a$\;
            $de_{ui} \gets E$\;
            \eIf{$RE < E$}
                {$W_i \gets w_u$\;
                 $RE \gets E$\;}
                {$W_i \gets W_i + w_u $\;}}
        \Return\;
        
        \function \claim{$u$}\\
        \updateState{$O, B, E, ES, RS$}\;
        $i \gets E~~mod~~2$\;
        \If{$de_{ui} \neq E - 1$ \textbf{or} $c = 0$ \textbf{or} $d_{ui} = 0$}{\Return\;}
        \eIf{$ce_u = E$}
            {\If{$cr_u = R$}
                {\Return\;}
            }
            {$ce_u \gets E$\;}
            $cr_u \gets R$\;
            $b_u \gets b_u + \min{(d_{ui}, s * w_u)}$\;
            $d_{ui} \gets d_{ui} - \min{(d_{ui}, s * w_u)}$\;
            $c \gets c - \min{(d_{ui}, s * w_u)}$\;
            \If{$d_{ui} = 0$}{$W_i \gets W_i - w_u$\;}
        \Return\;
        \end{algorithm}
}

%% file: heap_results.tex
\begin{table}[ht]
    \centering
    \begin{tabular}{rrr}
    Function & Present Study & Eth-heap\\
    \hline
    \hline
    Insert          &  95.459       & 101.261\\
    \hline
    Delete Minimum  & 133.272       & 170.448\\
    \hline
    \end{tabular}
    \caption{Average gas costs for \textit{Insert} and \textit{Delete Minimum} functions}
    \label{heap_results}
\end{table}

%% file: amf_performance.tex
\begin{table}[ht]
    \centering
    \scalebox{0.91}{
        \begin{tabular}{c|c|cc}
        Function    & No. of Users    & AMF & WAMF \\ 
        \hline  \hline
        \multirow{4}{*}{Demand} & $10$  & $70.245$ & $79.732$ \\ \cline{2-4}
                                & $50$  & $67.351$ & $77.135$ \\ \cline{2-4}
                                & $100$ & $66.989$ & $76.835$ \\ \cline{2-4}
                                & $500$ & $66.700$ & $71.365$ \\ \hline \hline
        \multirow{4}{*}{}       & $10$  & $46.800/140.401$ & $46,643/145.931$\\ \cline{2-4}
                    Claim       & $50$  & $42.240/126.720$ & $44.852/134.558$\\ \cline{2-4}
                (Avg./Total)    & $100$ & $42.114/126.344$ & $44.763/134.289$\\ \cline{2-4}
                                & $500$ & $42.047/126.143$ & $45.319/135.959$\\ \hline
        \end{tabular}
    }
    \caption{Average and total gas costs of AMF/WAMF \textit{demand} and \textit{claim} functions for various numbers of users.}
    \label{amf_performance}
\end{table}

%% file: amf_parameters.tex
\begin{table}[ht]
    \centering
    \begin{tabular}{rrp{3.5cm}}
    Parameter       & Value & Definition \\
    \hline
    \hline
    Number of Users & $n$     & The number of users in the system\\
    \hline
    Epoch Capacity  & $20n$  & The amount to be distributed for each epoch\\
    \hline
    Epoch Span      & $4n$   & The duration of an epoch in number of blocks\\
    \hline
    Round Span      & $n$     & The duration of a round in number of blocks\\
    \hline
    Demand Interval & $[10,30)$ & The interval from which the demands are drawn\\
    \hline
    \end{tabular}
    \caption{The values used in the tests for AMF and WAMF.}
    \label{amf_parameters}
\end{table}

%% file: claim.tex
\begin{table}[ht]
    \centering
    \begin{tabular}{rrr}
    \multicolumn{1}{c}{Round}   & \multicolumn{1}{c}{AMF}   & \multicolumn{1}{c}{WAMF}\\
    \hline
    \hline
    $1$ & $64.677$ & $67.211$\\
    \hline
    $2$ & $32.717$  & $36.158$\\
    \hline
    $3$ & $28.749$  & $32.589$\\
    \hline
    Average & $42.047$  & $45.319$\\
    \hline
    Total & $126.143$  & $135.959$     \\
    \hline
    \end{tabular}
    \caption{The costs of the \textit{claim} functions over rounds, in a setting with $n=500$ users.}
    \label{claim}
\end{table}